\documentclass[twocolumn]{aastex62}
\received{\today}
\revised{}
\accepted{}
\submitjournal{ApJ}

\usepackage{amsmath}
\usepackage{graphicx}

\graphicspath{./fig/, ./png/}

\renewcommand{\bm}{\boldsymbol}

\newcommand{\bbb}{{\bm b}}

\newcommand{\eee}{{\bm e}}

\newcommand{\uuu}{{\bm u}}

\newcommand{\AAA}{{\bm A}}

\newcommand{\BBB}{{\bm B}}
\newcommand{\mBBB}{\overline{\bm B}}
\newcommand{\mB}{\overline{B}}

\newcommand{\mEEE}{\overline{\boldsymbol{\cal E}}}
\newcommand{\mEE}{\overline{\cal E}}

\newcommand{\UUU}{{\bm U}}
\newcommand{\mUUU}{\overline{\bm U}}
\newcommand{\mUU}{\overline{U}}

\newcommand{\aalpha}{\boldsymbol{\alpha}}
\newcommand{\bbeta}{\boldsymbol{\beta}}
\newcommand{\ggamma}{\boldsymbol{\gamma}}
\newcommand{\ddelta}{\boldsymbol{\delta}}
\newcommand{\kkappa}{\boldsymbol{\kappa}}
\newcommand{\Tr}{\operatorname{Tr}\!}
\newcommand{\rmid}{r^{\rm m}}
\newcommand{\Eq}[1]{Eq.~(\ref{#1})}

\newcommand{\EQ}{\begin{equation}}
\newcommand{\EN}{\end{equation}}
\newcommand{\EQA}{\begin{eqnarray}}
\newcommand{\ENA}{\end{eqnarray}}
\newcommand{\brac}[1]{\langle #1 \rangle}
\newcommand{\dert}[1]{\frac{\partial #1}{\partial t}}

\newcommand{\trms}{{\rm rms}}
\newcommand{\urms}{u_{\rm rms}}

\newcommand{\ku}{k_u}

\newcommand{\Co}{{\rm Co}}

\newcommand{\Pra}{{\rm Pr}}
\newcommand{\PraSGS}{{\rm Pr}_{\rm SGS}}

\newcommand{\PrM}{{\rm Pr}_{\rm M}}

\newcommand{\Rey}{{\rm Re}}

\newcommand{\ReM}{{\rm Re}_{\rm M}}

\newcommand{\Ta}{{\rm Ta}}

\newcommand{\nab}{\mbox{\boldmath $\nabla$} {}}

{}

\newcommand{\DR}{DR }  
\newcommand{\Fig}[1]{Figure~\ref{#1}} 
\newcommand{\Figu}[1]{Figure~\ref{#1}}

\newcommand{\Table}[1]{Table~\ref{#1}}
\newcommand{\App}[1]{Appendix~\ref{#1}}

\shorttitle{Stellar dynamos in the transition regime}
\shortauthors{Viviani et al.}

\begin{document}

\title{Stellar dynamos in the transition regime: multiple dynamo 
modes and anti-solar differential rotation}

\correspondingauthor{M. Viviani}
\email{viviani@mps.mpg.de}

\author{M. Viviani}
\affil{Max-Planck-Institut f\"ur Sonnensystemforschung,
              Justus-von-Liebig-Weg 3, D-37077 G\"ottingen, Germany}

\author{M. J. K\"{a}pyl\"{a}}
\affil{Max-Planck-Institut f\"ur Sonnensystemforschung,
              Justus-von-Liebig-Weg 3, D-37077 G\"ottingen, Germany}
\affiliation{ReSoLVE Centre of Excellence, Department of Computer Science, 
              Aalto University, PO Box 15400, FI-00076 Aalto, Finland}              

\author{J. Warnecke}
\affil{Max-Planck-Institut f\"ur Sonnensystemforschung,
              Justus-von-Liebig-Weg 3, D-37077 G\"ottingen, Germany}

\author{P. J. K\"{a}pyl\"{a}}
\affiliation{Institut f\"ur Astrophysik, Georg-August-Universit\"at
  G\"ottingen, Friedrich-Hund-Platz 1, D-37077 G\"ottingen, Germany}
\affiliation{ReSoLVE Centre of Excellence, Department of Computer Science, 
              Aalto University, PO Box 15400, FI-00076 Aalto, Finland}    

\author{M. Rheinhardt}
\affiliation{ReSoLVE Centre of Excellence, Department of Computer Science, 
              Aalto University, PO Box 15400, FI-00076 Aalto, Finland}   

\begin{abstract}

Global and semi-global convective dynamo simulations of solar-like
stars are known to show
a transition from an anti-solar (fast poles, slow equator) to
solar-like (fast equator, slow poles) differential rotation 
(DR) for increasing rotation rate. 
The dynamo solutions in the latter regime can exhibit
regular cyclic modes, whereas in the former one,
only stationary or temporally irregular solutions have been obtained so far.  
In this paper we present a semi-global dynamo simulation   
in the transition region, exhibiting two coexisting dynamo modes,
a cyclic and a stationary one, both being dynamically significant.
We 
seek to understand how such a dynamo is driven by analyzing
the large-scale flow properties (DR and
meridional circulation) together with the turbulent transport
coefficients obtained with the test-field method.
Neither an $\alpha\Omega$ dynamo wave nor 
an advection-dominated dynamo are able to explain the cycle period and the
propagation direction of the mean magnetic field.
Furthermore, we find that the $\alpha$ effect is comparable
or even larger than the $\Omega$ effect in generating the toroidal
magnetic field, and therefore, the dynamo seems to be
$\alpha^2\Omega$ or $\alpha^2$ type.
We further find that the effective large-scale flows are significantly 
altered by turbulent pumping.
\end{abstract}


\section{Introduction} \label{sec:intro}

Recently, \cite{2018ApJ...855L..22B}
reported 
on
an abrupt increase of the magnetic activity level 
of solar-like stars
with decreasing values of the Coriolis number in the vicinity of its
solar value, with the Coriolis number quantifying the rotational
influence on convection.
Another observational study \citep{Olspert18} found that 
the degree of magnetic variability abruptly decreased, 
indicative of the disappearance of magnetic cycles, 
at slightly lower than solar chromospheric activity index values.
Moreover, \cite{2016ApJ...826L...2M} interpreted 
\textit{Kepler}
data
to indicate that the Sun is rotationally and magnetically in
a transitional state, where the global magnetic dynamo is 
shutting down.
\cite{2018ApJ...855L..22B} proposed
a transition in the differential rotation (DR)
from solar-like (for younger stars) to anti-solar  
(at a later age) to be responsible 
for some of these phenomena.

This transition (henceforth AS-S transition)
has already been the subject
of many numerical studies
\citep[see, e.g.,][]{GYMRW14, KKB14, MMK15, FM15,VWKKOCLB18}
and they all pinpoint it in a narrow
Coriolis number interval around its solar value. 
The latter can be estimated, for example, from mixing-length
models to be around two \citep[e.g.][]{KKB14}.
However, none of these works considered dynamo solutions 
near the transition point. They either studied the cyclic 
modes in the solar-like rotation regime, or the stationary 
and temporally irregular ones \citep{KKKBOP15, Warnecke2018}
obtained in the anti-solar regime.

In a previous paper \citep{VWKKOCLB18}, we
 reported on dynamo simulations of solar-like stars with
 varying rotation rate, two of which showed oscillatory behavior   
 in the AS-S transition.
In these simulations, the poleward migration of the magnetic field is
accompanied by a rotation profile exhibiting a decelerated equator
and faster polar regions (anti-solar DR).
The aim of the present paper is to study how 
such transitional--regime dynamos operate.

In the regime of solar-like DR,
cyclic dynamo solutions
with equatorward dynamo waves are often obtained from global
magneto-convection models \citep[e.g.][]{KMB12,ABMT15,SBCBN17}.
Most of them can be explained in terms of Parker waves 
\citep[see, e.g.,][]{WKKB14,WKKB16,WRTKKB18,KKOBWKP16,KKOWB16,Warnecke2018}.
The migration direction and cycle period of 
such waves is   
determined by the product of the $\alpha$ effect and the 
radial gradient of the local rotation rate $\Omega$ \citep{Pa55a, Yo75}.
For an equatorward-migrating field in the northern hemisphere (as observed on
the Sun), one needs, for example, a
negative radial gradient of $\Omega$ and a positive $\alpha$ effect.
However, simplified dynamo models often invoke an advection-dominated
concept \citep[e.g.][]{CSD95,DC99,KRS01} to explain 
the migration and cyclic behavior of 
large-scale stellar magnetic fields.
In this case, the meridional flow speed and direction at the location of the toroidal field
generation determine the cycle period and 
latitudinal dynamo wave direction.

Another possible mechanism generating cyclic dynamo solutions is an
$\alpha^2$ dynamo \citep{BS87,Ra87,Br17}. 
In this case, magnetic field generation is due to
the $\alpha$ effect alone,
and DR is not needed.
Such a dynamo was reproduced in forced turbulence
in a spherical shell \citep{MTKB10} and convection simulations
in a box \citep{MS14}, but global convective dynamo models 
have not yet yielded a similar solution.

In this work, we will 
investigate the properties of one particular transitional--regime 
dynamo solution, and test which
mechanisms can explain the seen cyclic behavior.
To achieve this goal we will use the test-field method 
\citep{SRSRC05,SRSRC07} 
for extracting
the turbulent transport coefficients.
This is possible due to the dominance of the axisymmetric magnetic field 
allowing us to try a description in terms of mean-field theory.
The test-field method has been 
successfully used in previous studies
to explain
planetary dynamos \citep[e.g.][]{Sch11,SPD11,SPD12}, 
cyclic dynamo solutions 
of solar-type stars
\citep{WRTKKB18,Warnecke2018}, and 
the long-term
variations of these solutions
\citep{GKW17}.

\section{Setup and Methods} \label{sec:setup}

We use the {\sc Pencil Code}\footnote{\url{https://github.com/pencil-code/}} to
solve the fully compressible magnetohydrodynamic equations
for the velocity $\UUU$,  the density $\rho$, the specific entropy $s$
and the magnetic vector potential $\AAA$ with the magnetic field
 $\BBB=\nab\times\AAA$ in a spherical shell without polar cap, defined
in spherical coordinates $(r,\theta,\phi)$ by
$0.7R \leq r \leq R$ for the radial extent, with
$\theta_0\leq\theta\leq\pi-\theta_0$ and $0\leq\phi\leq 2\pi$ 
for the extents in colatitude and longitude, respectively, 
where $\theta_0=15^{\circ}$. 
The setup is the same as the one used in \cite{KMCWB13} and \cite{VWKKOCLB18}.
We impose impenetrable and 
stress-free
boundary conditions
at all radial and latitudinal boundaries for the velocity field $\UUU$,  
and a perfect-conductor boundary condition at the bottom 
and the latitudinal boundaries for $\BBB$, while at the top, the field is 
forced to be radial.
The temperature follows a 
blackbody
condition at the top, whereas
a constant heat flux is prescribed at the bottom.
At the latitudinal boundaries, zero heat flux is enforced.
We start with an isentropic atmosphere for density and entropy, 
see \cite{KMCWB13} for details.
The initial conditions for the magnetic field and the velocity 
are weak Gaussian seeds.

Nondimensional
input parameters for the examined run are
the Taylor number, or correspondingly the Ekman number, defined as
\begin{equation}\label{TaylorNum}
  \Ta = \left( 2\Omega_0 \left( \Delta r\right)^2 / \nu \right)^2 = {\rm Ek}^{-2}=2.03 \cdot 10^{7},
\end{equation}
where $\Omega_0$ is the overall rotation rate with
$\Omega_0/\Omega_{\sun}=1.8$ for the considered run,
$\Delta r = 0.3 R$ is the thickness of the shell, and $\nu$ is the constant viscosity.    
Further, we have the thermal, 
sub-grid scale (SGS) 
thermal and magnetic Prandtl numbers,
the latter two describing
the unresolved turbulent effects:
\begin{equation}\label{PrNum}
\Pra\!=\!\frac{\nu}{\chi^{\rm m}} = 58, \
\PraSGS\!=\!\frac{\nu}{\chi^{\rm m}_{\rm SGS}} = 2.5, \
\PrM\!=\!\frac{\nu}{\eta} = 1,
\end{equation}
Here,
$\chi^{\rm m}$ is the heat diffusivity
calculated in the middle of the convective zone
at $\rmid=0.85R$ as 
$\chi^{\rm m} = K\left(\rmid \right)/c_P \rho \left(\rmid\right) $,
$c_P$ being the specific heat at constant pressure. The radiative heat
conductivity $K$ follows an $r^{-15}$ dependency to mimic the actual
heat flux profile in the Sun.
$\chi_{\rm SGS}^{\rm m}$ is the turbulent heat  
diffusivity at $r=\rmid$
\citep[see][for details]{KMCWB13}
and $\eta$
is the constant magnetic diffusivity.

The non-dimensional quantities are scaled to physical units using
the solar radius $R = 7\cdot10^{8}~{\rm m}$,
solar rotation rate $\Omega_{\sun}=2.7\cdot10^{-6}~{\rm s^{-1}}$, 
the density at the bottom of the solar convection zone 
$\rho(0.7R)=\rho_0=200~{\rm kgm^{-3}}$,
and $\mu_0=4\pi\cdot10^{-7}~{\rm Hm^{-1}}$.  
The initial density contrast in the simulation is roughly 
30, and the dimensionless luminosity $\mathcal{L}=L_0/[\rho_0     
  (GM)^{3/2}R^{1/2}]\approx3.8\cdot10^{-5}$, where $L_0$ is the
luminosity in the simulation, $G$ is the
gravitational constant and $M$ the mass of the star. This 
corresponds to an approximately
$10^6$ times higher luminosity than 
the solar one,
$L_\odot$,
 to avoid the acoustic timestep constraint. The rotation rate is
increased correspondingly in proportion to 
$\left(L_0/L_\odot\right)^{1/3}$,
 to obtain a realistic rotational influence on the
flow \citep[for further details see Appendix~A of][]{KGOKB19}.

We indicate by $\mBBB$ and $\mUUU$ the mean
(longitudinally averaged) fields, and by $\bbb'$, $\uuu'$ the
corresponding fluctuating fields, so that, for example, ${\BBB}=\mBBB+\bbb'$.

The need to compute turbulent transport coefficients can be seen from
the induction equation for the mean magnetic field, $\mBBB$:
\begin{equation}\label{eq:meanB}
\dert{\mBBB}=\nab\times\left(\mUUU\times\mBBB+\overline{\uuu'\times\bbb'}\right)-\nab\times\eta\nab\times\mBBB.
\end{equation}
The term $\mEEE=\overline{\uuu'\times\bbb'}$ is the turbulent 
electromotive force
(EMF);
it can be expanded in terms of $\mBBB$ and its derivatives.
Further,
the tensorial coefficients of the individual
contributions can be divided into symmetric and anti-symmetric parts
\citep[see, e.g.,][]{KR80} such that
\begin{equation}\label{eq:emf}
\mEEE=\aalpha\cdot\mBBB+\,\ggamma\times\mBBB
-\bbeta\cdot\nab\times\mBBB-\ddelta\times\nab\times\mBBB
-\kkappa\cdot\left(\nab\mBBB\right)^{(s)}\!\!\!,
\end{equation}
where $\aalpha$ and $\bbeta$ are symmetric tensors of rank two,   
$\ggamma$ and $\ddelta$ are vectors,
while
$\kkappa$ is a tensor of rank three  
with $\left(\nab\mBBB\right)^{(s)}$ being the symmetric 
part of the derivative tensor of $\mBBB$.
Each of these coefficients can be related to a physical effect,
e.g., 
$\aalpha$ covers cyclonic generation ($\alpha$ effect),  
$\bbeta$ describes turbulent diffusion,
$\ggamma$ represents turbulent pumping.
The pumping enters the {\it effective mean flow},
$\mUUU^{\rm eff}\!\!=\mUUU+\ggamma$,
 \citep[e.g.][]{Ki91,OSBG02,KKT06,WRTKKB18}
and may thus be crucial in determining the nature of the dynamo.

To determine the turbulent transport coefficients, 
we continued 
one of the transitional--regime dynamo runs
from \cite{VWKKOCLB18}, showing
a cyclic dynamo solution (Run~C1), 
with the test-field module of the {\sc Pencil Code}  
activated \citep[for its theory, see][]{SRSRC05,SRSRC07}.  

\section{Results} \label{sec:results}

The run considered is characterized   
by the following nondimensional output parameters:
the fluid and magnetic Reynolds numbers
\begin{equation}
\Rey = \frac{\urms}{\nu \ku} = 41, \quad
\ReM = \frac{\urms}{\eta \ku} = 41, \quad
\end{equation}
and the Coriolis number
\begin{equation}
\Co = \frac{2\Omega_0}{\urms \ku} = 2.8.
\end{equation}
Here,
$\ku=2\pi/\Delta r\approx21/R$ 
is an estimate of the
wavenumber of the largest eddies,
and the   
averaged rms velocity is defined as
$\urms=\sqrt{(3/2)\brac{U_r^2+U_\theta^2}_{r\theta\phi t}}$
 \citep[see][]{KMCWB13}.
Angle brackets indicate averaging over the coordinate(s) in the subscript.

\subsection{Mean magnetic field}
The mean magnetic field is prevailingly symmetric about the equator 
(quadrupolar) and shows cyclic behavior with 
poleward-migrating
 $\mB_\phi$,    
and polarity reversals at mid to high latitudes (\Fig{fig:bres}a). 
Detailed inspection of the solution reveals the presence of 
a cyclic and a stationary constituent,
the latter being 2-2.5 times stronger (in rms values) than the former.
We interpret these as two different,
coexisting dynamo modes, $\langle\mBBB\rangle_t$
and $\mBBB^{\,\rm cyc}
\!=\mBBB-\langle\mBBB\rangle_t
$, respectively;
see \Fig{fig:bres}b-d for the toroidal component of 
$\mBBB^{\,\rm cyc}$ at two depths,
along with its dependence on radius 
and time at latitude $+50^\circ$ where 
it is strongest in rms value.
Its topology is similar throughout the convection zone, 
and the poleward migration is present at all depths.

\begin{figure}
\begin{center}
\includegraphics[width=\columnwidth]{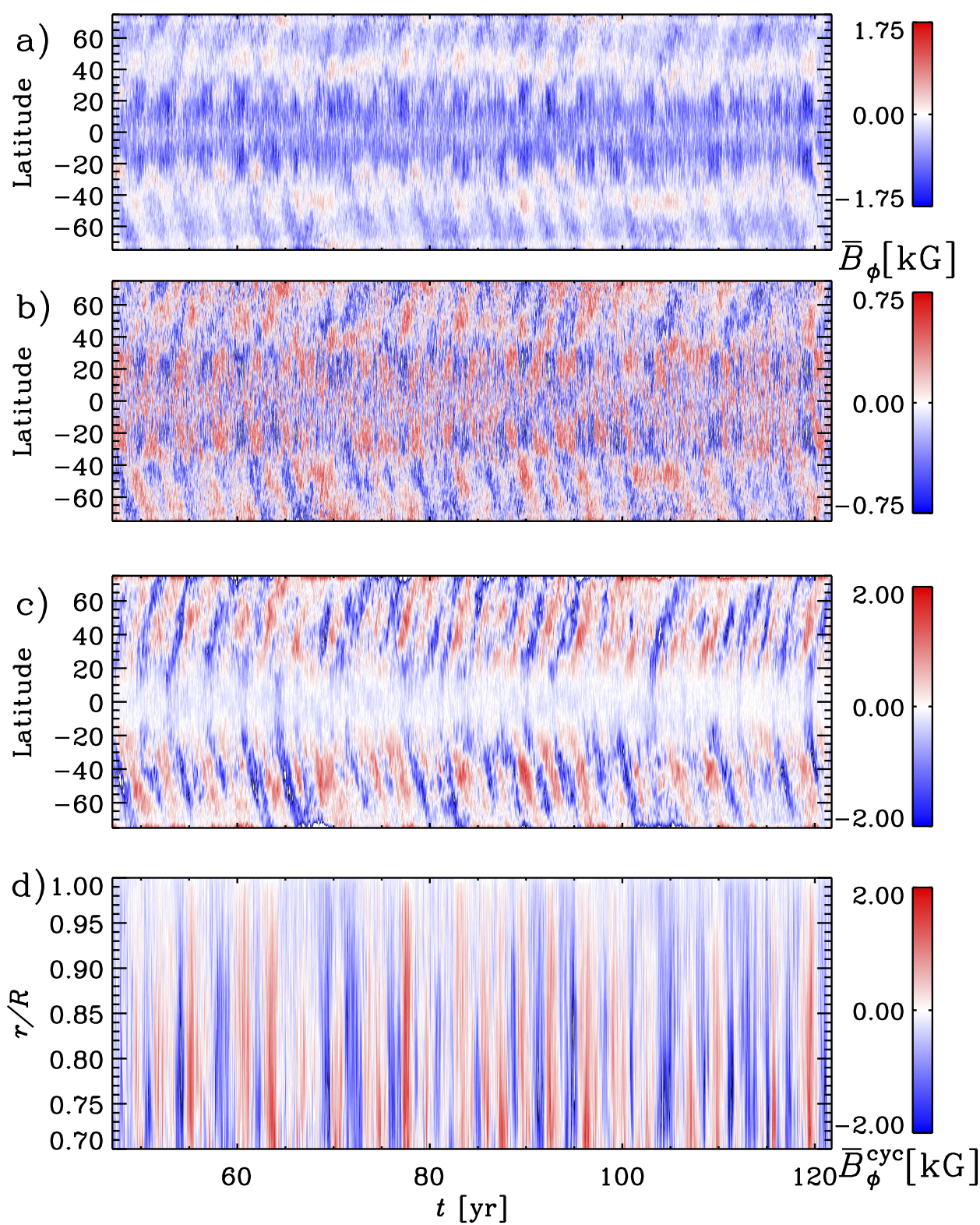}
\end{center}\caption{
(a): time-latitude diagram for $\mB_{\phi}$ near the surface
($r=0.98~R$).
(b): analogously for $\mB_{\phi}^{\,\rm cyc}=\mB_{\phi}-\langle\mB_{\phi}\rangle_t$.
(c): same as in (b),
but at $r=0.8~R$.
(d): time-radius diagram for $\mB_{\phi}^{\,\rm cyc}$ at latitude $50^\circ$. }
\end{figure}\label{fig:bres}

\begin{figure*}
\begin{center}
\includegraphics[width=2.\columnwidth]{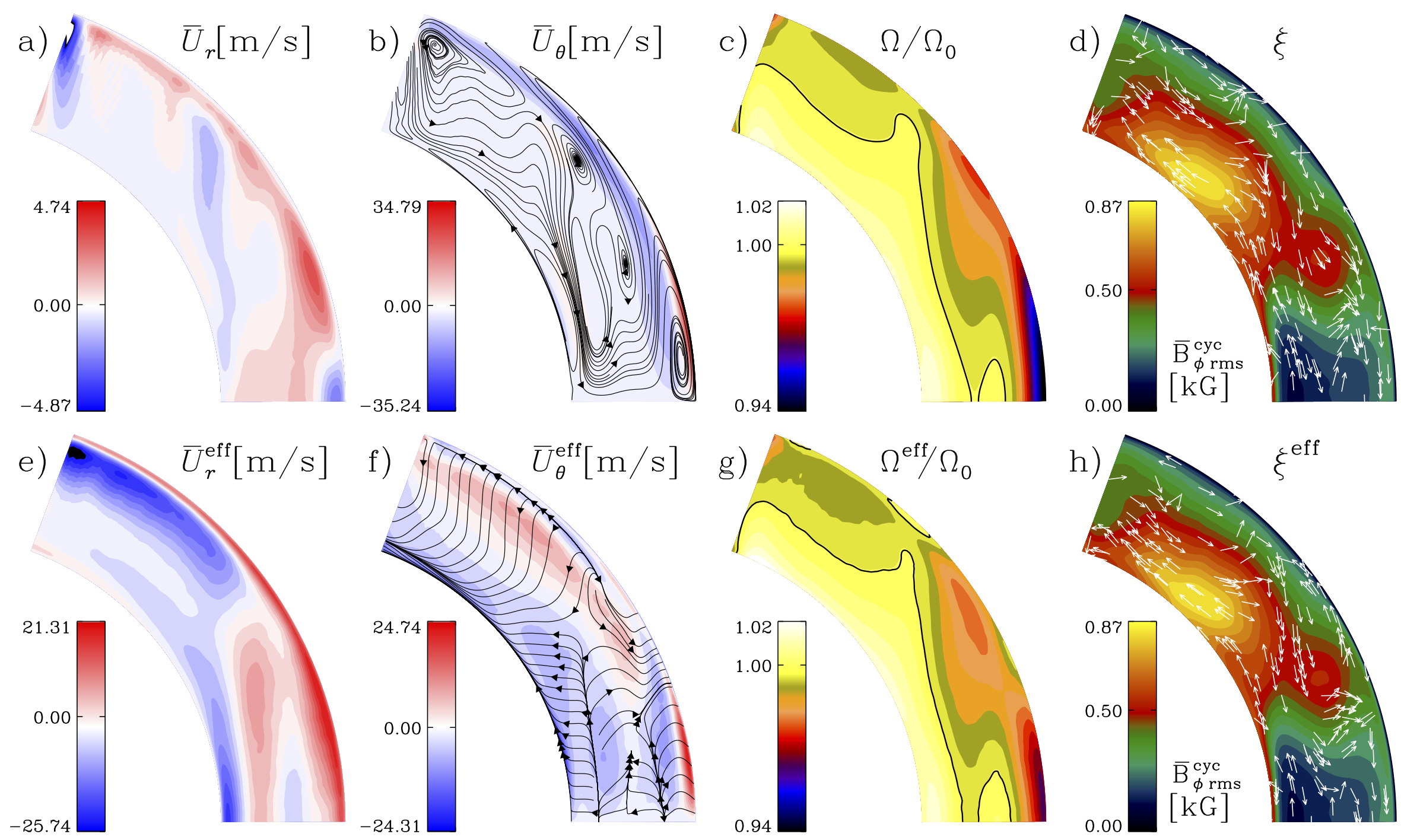}
\end{center}\caption{
Time-averaged radial (a) and latitudinal (b) components of the
meridional circulation ($\mUU_r ,\mUU_\theta$, 0), (c) mean angular velocity   
$\Omega=\mUU_\phi /r\sin\theta+\Omega_0$
and (d) temporal rms value of the 
azimuthal component of the cyclic
magnetic field
$\mB_\phi^{\,\rm cyc}$. 
(e)-(h): same as (a)-(d),
but using the effective mean velocity. 
Flow lines in (b), (f): 
meridional and effective meridional circulation, respectively.
Black lines in (c) and (g): $\Omega/\Omega_0=1$ and
$\Omega^{\rm eff}/\Omega_0=1$, respectively.
Arrows in d), h): direction of the Parker-Yoshimura dynamo wave   
propagation, see \Eq{eq:PYdir}.
}   
\end{figure*}\label{fig:eff_flow}

\subsection{Mean flows}
We start our analysis by investigating 
meridional circulation and DR
as shown in \Fig{fig:eff_flow}(a)-(c).
The former has a dominant, large, 
anticlockwise cell, 
producing a relatively strong (20 m\,s$^{-1}$) poleward flow 
near the surface at almost all latitudes.      
There is a slow equatorward return flow 
widely distributed in the bulk of the convection zone
at mid to high latitudes.
In the slow rotation regime, anti-solar \DR is
often accompanied by a single cell anti-clockwise meridional
circulation.
In contrast, in the 
regime of fast rotation,
solar-like \DR
drives multicellular 
meridional circulation aligned with the
rotation axis.
The cell pattern in this run represents a transitional state between
these two regimes \citep[e.g.,][]{KKB14,KKKBOP15,FM15}.

The DR
profile shows a decelerated equator and 
accelerated polar regions at the surface;
hence it is broadly speaking anti-solar, 
despite some regions of weakly solar-like \DR at the bottom of the CZ.
The pole-equator difference at
the surface is comparable to runs with similar rotational influence    
\citep[e.g.][]{KKKBOP15,Warnecke2018}.
However, the energy in the DR
compared to the total
kinetic energy, neglecting the rigid rotation, is smaller
than in runs with slightly slower 
and faster rotation
\citep{VWKKOCLB18}.
This is most likely because our run is very close to the 
actual AS-S transition.
In the DR profile, we find two distinct features:
at 
mid-latitudes there is a local minimum of $\Omega$,
which has also been
found in simulations with about three times faster rotation. 
In these,
 the resulting shear drives 
a dynamo wave obeying the Parker-Yoshimura rule
\citep[e.g.][]{WKKB14}.
Furthermore, we find strong negative shear in a layer near the surface at
low latitudes.

\begin{figure}
\begin{center}
\includegraphics[width=\columnwidth]{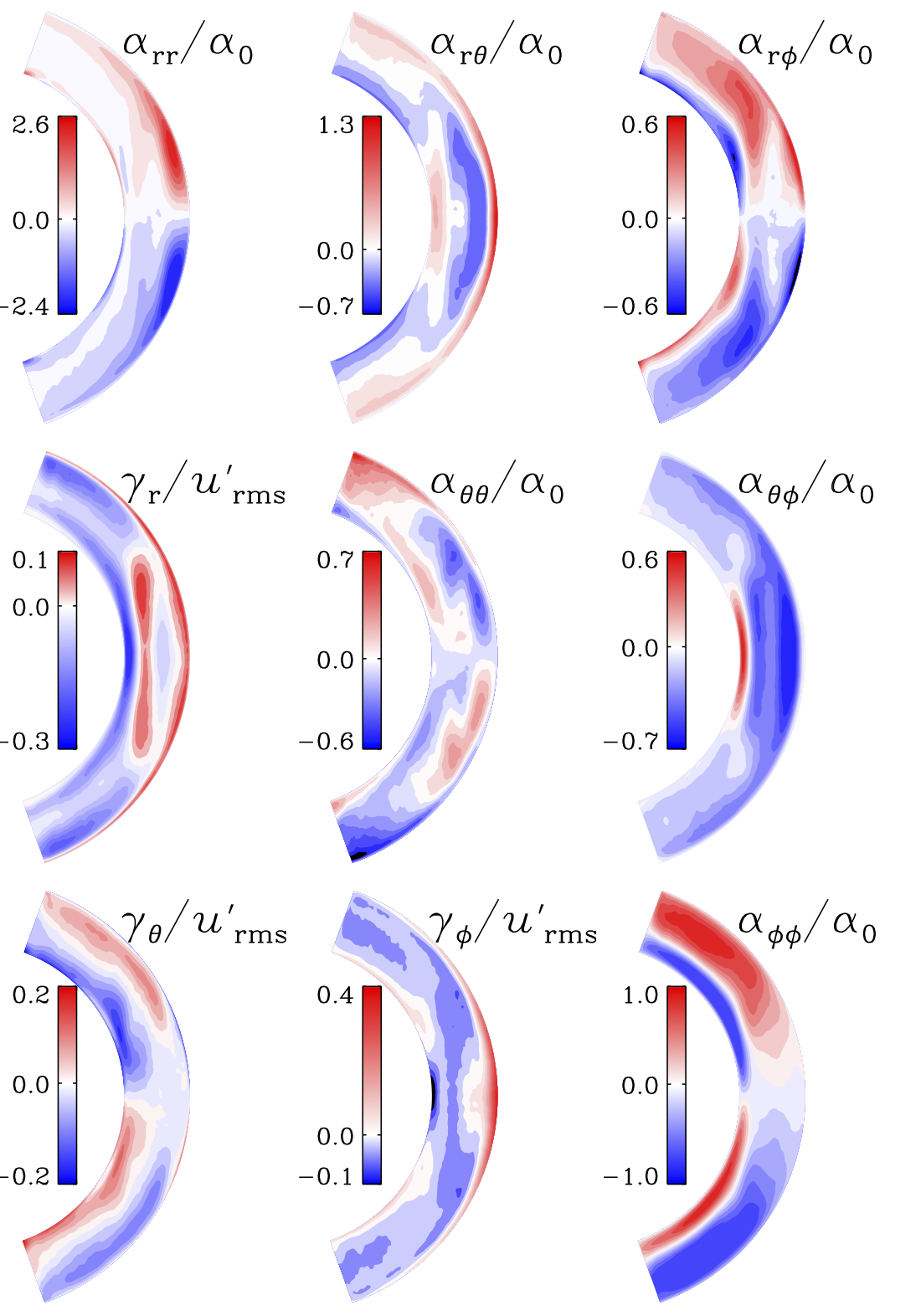}
\end{center}\caption{
Independent components of time-averaged
$\aalpha$, 
normalized by 
$\alpha_0=u'_{\trms}/3$,
and $\ggamma$, normalized by $u'_{\trms}$.
}
\end{figure}\label{fig:alpC1}

\subsection{Turbulent transport coefficients}

Next, we look at the turbulent transport coefficients, 
for which we have used a slightly different definition
than in previous work 
\citep[][]{SRSRC07,WRTKKB18}, see \App{sec:new_deco} for
motivation and details.
We begin by discussing
$\aalpha$ and $\ggamma$, and compare them with
their counterparts
from a more rapidly rotating dynamo run with
solar-like DR of \cite{WRTKKB18} in terms of the 
ratio of their extremal values.
Regarding $\aalpha$ (see \Fig{fig:alpC1}),
both $\alpha_{rr}$ and  $\alpha_{\phi\phi}$ are 20 \% smaller
than in \cite{WRTKKB18}, while the meridional profiles are similar.
Furthermore, $\alpha_{\theta \theta}$ is nearly 30\% larger and shows
an opposite sign near the surface close to the equator.
The corresponding ratios for the off-diagonal components
$\alpha_{r\theta}$, $\alpha_{\theta\phi}$, and $\alpha_{r\phi}$ are
1.9, 1.1, and 0.7, respectively. 
Moreover, $\alpha_{r\theta}$ and
$\alpha_{r \phi}$
show opposite signs at the equator near the surface.
We associate these differences from \cite{WRTKKB18} 
with the milder rotational influence on convection, characterized by 
the Coriolis number, being roughly three times
smaller in our run.
The usage of the new definition of the turbulent transport coefficients
could also have caused some of these differences,
but this influence was checked to be very small by re-computing
the coefficients for \cite{WRTKKB18} using the new convention.
A detailed comparison is shown in \Table{tab:coeff} of \App{sec:new_deco}.

Concerning the turbulent pumping
(see \Fig{fig:alpC1}), $\gamma_r$ 
has 
a similar magnitude,
$\gamma_{\theta}$ is 40\% weaker and
$\gamma_{\phi}$ is 40\% stronger
than in \cite{WRTKKB18}.
Here, too,
 the new definition has no significant effect.
Note also the different normalization we used for $\ggamma$.
$\gamma_r$ is upward everywhere 
except in the bulk of the convection zone at mid and
high latitudes.
$\gamma_\theta$ is equatorward (poleward) in the upper (lower) half of
the convection zone.
$\gamma_{\phi}$ is prograde near the surface and at mid-latitudes near
the bottom, and negative everywhere else.
The magnitudes of all three components are around 0.3 $u'_{\trms}$,
where $u'_{\trms}(r,\theta)=\langle{\boldsymbol{u}'}^2\rangle^{1/2}_{\phi t}$ is the local   
turbulent rms velocity in the meridional plane.
The effective mean velocity resulting from $\ggamma$
is shown by its time average
in \Fig{fig:eff_flow}(e)-(g).
The radial component, $\mUU_r^{\rm eff}$,
is completely dominated by $\gamma_r$, 
leaving nearly no trace of the
actual flow. 
$\gamma_{\theta}$ changes the sign of $\mUU_{\theta}$ only slightly below
the surface and reduces its magnitude by around 30\%. 
However, the meridional flows cells are completely destroyed, as shown by the flow
lines in \Fig{fig:eff_flow}f.
$\gamma_{\phi}$ is accelerating the equator and decelerating the polar region.
The larger change in $\Omega^{\rm eff}$ compared to
\cite{WRTKKB18}~is because $\gamma_{\phi}$ increases
with decreasing $\Omega_0$.

The reconstruction of the turbulent 
EMF $\mEEE$ 
based on \Eq{eq:emf} shows reasonable agreement with
$\overline{\uuu'\times\bbb'}$, see \App{sec:EMFrec}.
Therefore, we can confidently use the set of turbulent transport
coefficients to describe the dynamo processes in this run.

\subsection{Dynamo cycles and migration}\label{sec:dyncyc}
As a first step in determining the possible dynamo mechanism, we
compare the period of the magnetic field cycle with theoretical expectations.
We compute the magnetic cycle period by Fourier transforming
$\mB_{\phi}$ at $r=0.98R$ and then averaging the spectra over
latitude.
As a result, we get $P_{\rm cyc}=(3.2\pm 0.3)~{\rm yrs}$,
where the error is obtained from the width at half maximum.

The two main dynamo scenarios both make predictions for the dynamo
cycle length  $P_{\rm cyc}$. 
The Parker-Yoshimura dynamo period is locally
 defined as \citep{Pa55a, Yo75}
\begin{equation}\label{eq:PPY}
P_{\rm PY} = 2 \pi\left|\frac{\alpha_{\phi
    \phi}k_{\theta}}{2}r\operatorname{cos}\theta\,\partial_r    
    \Omega\right|^{-1/2},
\end{equation}
where $k_{\theta}=2\pi/(r\Delta\theta)$ is the latitudinal wavenumber
of the dynamo wave with
$\Delta\theta = \pi/2-\theta_0$.
The justification of using only $\alpha_{\phi\phi}$ in
Eq.~(\ref{eq:PPY}) is that the other contributions to the poloidal field
generation are smaller.

The cycle period of an advection-dominated 
dynamo is related to the travel time of the meridional circulation
from the equator to the pole, 
$\tau_{\rm MC}$, such 
that $P_{\rm MC}\approx 2\,\tau_{\rm MC}$
\citep{KRS01,KROS18}.
Hence, in our notations, the expected cycle period can be written as
\begin{equation}\label{eq:PMC}
P_{\rm MC}=\frac{2r\Delta\theta}{\mUU_{\rm MC}(r,\theta)}
\end{equation}
where $\mUU_{\rm MC}$ is the temporal rms
\footnote{We define the temporal rms for a quantity $f$ as $\sqrt{\langle f^2 \rangle_t} $.}
of the meridional flow at the 
location of the dynamo wave. 
Traditionally, advection-dominated 
dynamo models
assume the meridional flow and the resulting migration to be significant
near the bottom of the convection zone, which would correspond to
setting $r=0.7~R$, but in the present case 
it is not so straightforward to determine the location of the dynamo
wave.

We start by using the measured
radial DR in Eq.~(\ref{eq:PPY}),   
and meridional flow in Eq.~(\ref{eq:PMC}), and obtain for the averages
over the convection zone
$\langle P_{\rm PY}\rangle_{r\theta}=2.2~{\rm yr}$
and $\langle P_{\rm MC}\rangle_{r\theta}=8.2~{\rm yr}$.
Using the meridional circulation in the lower quarter of the convection zone
only, we obtain, instead, 
 $\left< P_{\rm MC} \right>_{\delta r \theta}=63.8~{\rm yr}$, 
 where $\delta r$ goes from $0.7R$ to $R/4$.
Considering the relevant role of the turbulent pumping, especially in 
$\mUU_r$,
we also calculated the periods using the effective velocity, 
that is adding the contributions of turbulent pumping to the measured
large-scale flows,
obtaining
$\langle P_{\rm PY}^{\rm eff}\rangle_{r\theta}=2.0~{\rm yr}$, 
$\langle P_{\rm MC}^{\rm eff}\rangle_{r\theta}=5.6~{\rm yr}$ 
and 
$\left< P^{\rm eff}_{\rm MC} \right>_{\delta r \theta}=22.0~{\rm yr}$. 
The Parker-Yoshimura periods
are less affected than those from meridional circulation, as the 
$\ggamma$
contribution is more significant for the meridional circulation
than for the DR.
In conclusion, the Parker-Yoshimura periods are 
consistent with the measured magnetic cycle, while advection by
meridional flow cannot explain it.

If the mean magnetic field was advected
by the meridional flow or its effective counterpart,
one would not be able to explain
poleward migration
virtually everywhere within the convection zone.
This becomes evident from 
\Fig{fig:eff_flow}(b) and (f), 
where
equatorward flows are present.
Whether the meridional circulation is able to overcome diffusion,
can be assessed by help of the
corresponding dynamo number  
(or turbulent magnetic Reynolds number)
\begin{equation}
C_U =\Delta r\,\langle \mUU_{\rm
  MC}\rangle_{r \theta} /\langle {\rm
  Tr\!\left\lbrace\bbeta\right\rbrace}\rangle_{r \theta},
\end{equation}
where $\Tr\,\{\cdot\}$ indicates the trace.
The time-averaged values for the measured mean
and the effective mean flow are 0.2 and 0.6, respectively. 
Values below unity imply that the (effective) flow cannot 
overcome diffusion, not even with $\ggamma$ included;
therefore, the advection-dominated dynamo scenario is not applicable here.
However, the obtained values indicate that the meridional circulation 
may not be completely negligible in the magnetic evolution.

The prediction for the Parker-Yoshimura wave propagation 
direction given by \citep{Yo75}
\begin{equation}\label{eq:PYdir}
\boldsymbol{\xi}(r, \theta)=-\alpha_{\phi\phi}\hat{\eee}_{\phi}\times\nab{\Omega},
\end{equation}
is depicted in \Fig{fig:eff_flow}d) and h)
for the shear from $\Omega$ 
and $\Omega^{\rm eff}$, respectively.
Near the bottom of the convection zone, where also the
cyclic field is strongest, $\boldsymbol{\xi}$
is poleward at almost all latitudes, which
would agree with the actual field propagation.
In the bulk of the convection zone, however, the predicted direction
is equatorward, failing to explain the actual migration.
Hence, neither the Parker-Yoshimura-rule-obeying dynamo wave
nor the
advection dominated dynamo alone can be responsible for the 
oscillating magnetic field, poleward migrating
throughout the convection zone.

\begin{figure*} 
\begin{center}
\includegraphics[width=2.0\columnwidth]{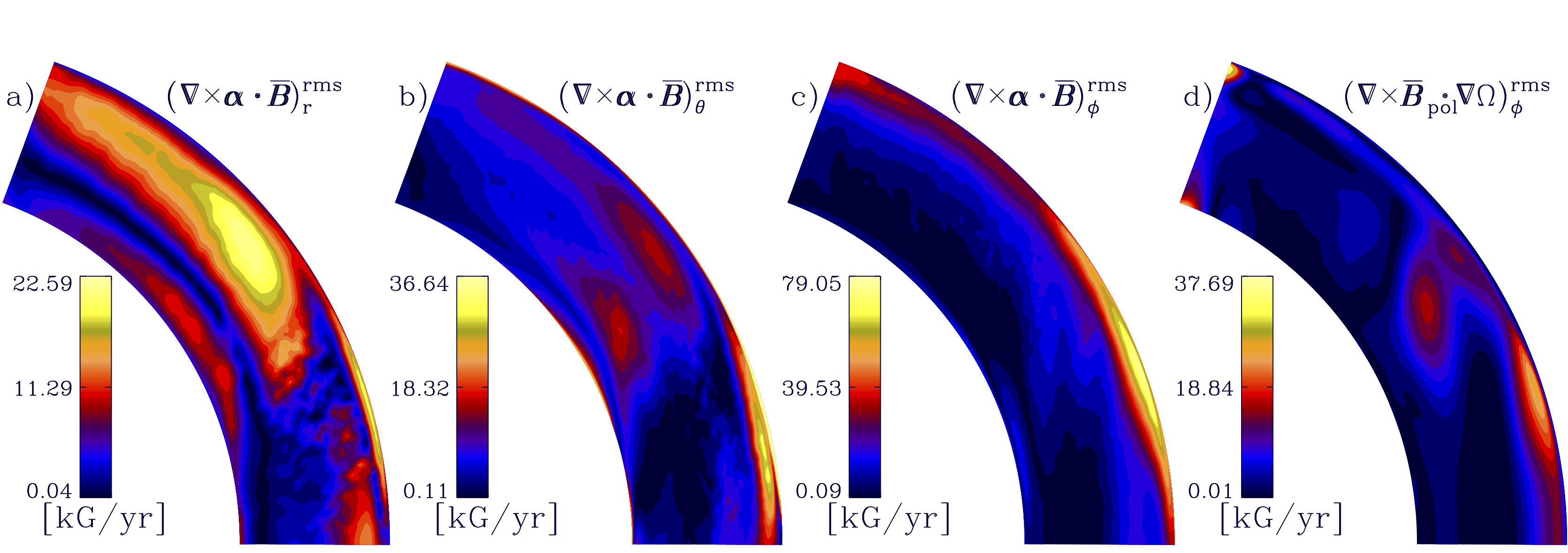}
\includegraphics[width=1.5\columnwidth]{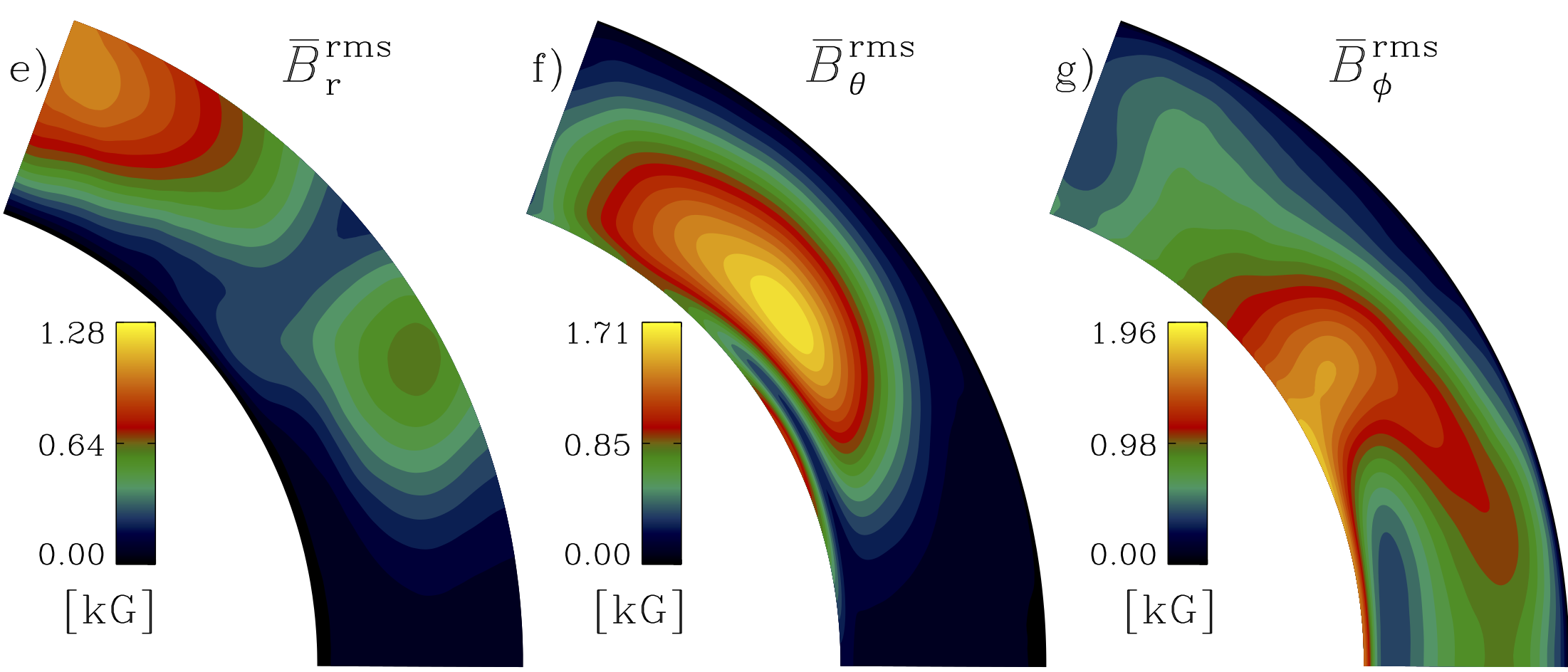}    
\end{center}\caption{
(a)-(d): temporal rms of the 
components of 
the $\alpha$ 
and $\Omega$ effect terms.
(a),(b): poloidal field generators;
(c),(d): toroidal field generators.
(e)-(g): temporal rms values of the components of $\mBBB$.
}
\end{figure*}\label{fig:effects}

\subsection{Dynamo drivers}
To understand the failure of the simple dynamo scenarios 
in explaining cycles and 
migration of the field, we finally turn to computing the terms
contributing to the magnetic field generation
in detail.
We present the contributions of the $\Omega$ and $\alpha$ effects, 
that is, of $\mBBB\cdot\nab\Omega$ and $\nab\times(\aalpha\cdot\mBBB)$, 
in terms of their temporal rms values
in \Fig{fig:effects} employing the total magnetic field (upper row), 
and show the corresponding temporal rms
magnetic fields in the lower row.
The two leftmost (rightmost) columns show the generators 
for the poloidal (toroidal) magnetic field.
From the magnitudes of the toroidal generators, it is evident that 
the $\alpha$ effect is equally important, or even dominant 
over the $\Omega$ effect. 
Hence, the generation of the toroidal field by the $\alpha$ effect 
is more efficient than  by the $\Omega$ effect,
suggesting an $\alpha^2\Omega$ or even an $\alpha^2$ dynamo mechanism
for the observed dynamo.

The $\Omega$ effect generates toroidal field efficiently at 
low latitudes near the surface and at mid-latitudes in the bulk of the convection zone, 
coinciding with the 
surroundings of the local minimum of $\Omega$.
The $\alpha$ effect is strongest near the surface, 
but shows also toroidal field generation 
around the local minimum of $\Omega$.
The patches of strong
rms toroidal field, however, overlap only partially with its generators,
and its profile is clearly offset deeper into the convection zone. 
One reason might be the radial-field boundary condition, which 
suppresses any toroidal field near the surface.
The $\alpha$ effect generates poloidal field mostly at high latitudes at all depths of the convection zone,
although there are also regions of strong field generation close to the surface near the equator.
The high-latitude field generator profiles match qualitatively better to the rms
poloidal field distribution than to that of
the toroidal field, but still the match is very incomplete.

The mismatch between the generators and the actual field distribution 
indicates that our conclusion of the generating
mechanism being a simple $\alpha^2\Omega$ or $\alpha^2$ dynamo is
not a very solid one, and
that other dynamo effects might be at play.
For example, we find that the $\delta$ (R\"adler) effect 
may also redound to the driving of the dynamo.
Its contribution,
shown in \Fig{fig:deleff} in \App{sec:new_deco},
is significant 
near the surface, at mid-latitudes
for the poloidal field (panels a-b) and at all latitudes
for the toroidal field (panel c).
Particularly in the latter case, the contribution of $\ddelta$ is  
strong in the same regions as the $\alpha$ and $\Omega$ effects
and with roughly the same magnitude.
However, this effect, in its simplest form in a shear flow, is known to 
lead to stationary solutions \citep{BS05}.
Hence, its role for the oscillatory dynamo mode is likely to be negligible. 
How the $\delta$ effect contributes
to the magnetic field generation
needs to be analyzed in detail using mean-field simulations.

The study of \cite{Warnecke2018} covers parameter 
regimes very close to the one explored here, but all of these 
solutions appear to exhibit only stationary or temporally irregular modes.
This draws attention to the role of the wedge assumption
used in that study.
There, the computational domain covers only $\pi/2$ in azimuth, instead
of the full $2\pi$ interval here, being virtually the only difference
between these two studies.
Our interpretation is that there are various dynamo modes excited    
with very similar critical dynamo numbers. 
In terms of dynamo theory, the coexistence of a steady and an
oscillating field constituent can be understood as follows:
sufficiently overcritical flows enable the growth of more than one dynamo mode.
Under the assumption of steady mean flows and statistically stationary turbulence,
the time dependence of these
eigenmodes is exponential with an, in general, complex increment.
It is well conceivable that a non-oscillating and an oscillating mode are both excited 
and even continue to coexist in their nonlinear stage,
although their kinematic growth rates were different. 
 
\section{Conclusions} \label{sec:conc}

We presented and analyzed a spherical convective dynamo simulation 
located in the transitional regime between S and AS rotation profiles.
Unlike the oscillatory or stationary/irregular dynamos, of the S and AS regimes,
the dynamo consists of coexisting cyclic and stationary modes.
\cite{2016ApJ...826L...2M} suggested that the drop in
the variability level of stars slightly less active than the Sun could be the result 
of a shutdown of the dynamo.
Motivated by our finding of coexisting cyclic and stationary
modes, we rather interpret this drop
to be due to a change in the dynamo type.
We tried to explain the oscillating magnetic field as a 
Parker-Yoshimura-rule-obeying
dynamo wave or
within the 
advection-dominated 
framework.
Neither of the two approaches alone can explain the results in terms
of cycle period and migration direction, even if we take the turbulent
contributions to the effective mean flow into account.
One reason might be that the $\alpha$ effect plays 
here a more dominant role than in a simple $\alpha\Omega$ dynamo.
Our claim is validated by the analysis of the 
field generators shown in \Fig{fig:effects}:
the mean field is generated by cyclonic convection 
and DR together,
suggestive of an $\alpha^2\Omega$ or $\alpha^2$ dynamo.
However, the spatial distributions of the generators do not match very
well with those of the mean fields. 
This likely indicates that other  
dynamo effects may also play important roles, and we find evidence of a 
significant contribution from the $\delta$ effect.
However, mean-field models that take into account all turbulent effects
are needed to address this issue.

\appendix

\section{Redefinition of the turbulent transport coefficients}
\label{sec:new_deco}
\subsection{Motivation}
As mentioned in \cite{SRSRC07} and \cite{WRTKKB18}, there is some
arbitrariness in deriving the transport coefficients (see \Eq{eq:emf})
from the (non-covariant) tensors $\tilde{\boldsymbol{a}}$ and
$\tilde{\boldsymbol{b}}$ defined by 
\begin{equation}
  \mEE_\kappa = \tilde{a}_{\kappa\lambda} \mB_\lambda + \tilde{b}_{\kappa\lambda r} \partial_r \mB_\lambda +  \tilde{b}_{\kappa\lambda \theta} \partial_\theta \mB_\lambda,   
  \quad \kappa,\lambda = r,\theta,\phi
  \label{eq:noncovEMF}
\end{equation}
which form the immediate outcome of the test-field method. Here, we
specify a choice, different from the one employed earlier
\citep[see][]{SRSRC07,WRTKKB18,Warnecke2018}, and characterized
by a maximum of vanishing components in $\kkappa$.
As a consequence, the role of $\kkappa$
in the turbulent EMF
$\mEEE$ is reduced, while
mainly that of $\bbeta$ is enhanced. 
This is motivated by the difficulty to interpret $\kkappa$
physically, whereas $\bbeta$ clearly stands for turbulent dissipation.
As a meaningful side effect, the diagonal elements of the latter become 
equal for isotropic turbulence. 
Furthermore, localized appearances of negative definite 
$\bbeta$, which are destructive to mean-field modeling, 
become more visible as less of the diffusive contributions 
(ideally none) are ``hidden" in $\kkappa$.
Thus, removing the negative definiteness in the redefined 
$\bbeta$ has better prospects to render the mean-field model feasible.

\subsection{Decomposition}
In \Eq{eq:noncovEMF}, the components $\tilde{b}_{\kappa\lambda \phi}$ 
do not appear as all $\phi$ derivatives vanish. 
They show up in the definitions of $\aalpha$, $\bbeta$,
etc. though, but setting them arbitrarily cannot change $\mEEE$. 
Here, we choose $\tilde{b}_{\kappa\lambda \phi}=-\tilde{b}_{\kappa\phi\lambda}$, 
in contrast to \cite{SRSRC07} who set $\tilde{b}_{\kappa\lambda \phi}=0$. 
Then we arrive at the following  expressions for the transport coefficients,
where underlines indicate new or altered terms in comparison to   \cite{SRSRC07}:
\begin{eqnarray}
\alpha_{rr} &=& \tilde{a}_{rr}-\tilde{b}_{r\theta\theta}/r\\
\alpha_{r\theta}=\alpha_{\theta r} &=& \frac{1}{2}\big( \tilde{a}_{r\theta}+\tilde{a}_{\theta r}+(\tilde{b}_{rr\theta}-\tilde{b}_{\theta\theta\theta})/r\big)\\
\alpha_{r\phi}=\alpha_{\phi r} &=& \frac{1}{2} \big( \tilde{a}_{r\phi}+\tilde{a}_{\phi r} -(\underline{\tilde{b}_{r\phi r} + \cot \theta\, \tilde{b}_{r\phi\theta}}+\tilde{b}_{\phi\theta\theta})/r\big)\\
\alpha_{\theta\theta} &=& \tilde{a}_{\theta\theta}+\tilde{b}_{\theta r\theta}/r\\
\alpha_{\theta\phi}=\alpha_{\phi \theta} &=& \frac{1}{2}\big( \tilde{a}_{\theta\phi}+\tilde{a}_{\phi \theta} -(\underline{\tilde{b}_{\theta\phi r} + \cot \theta\, \tilde{b}_{\theta\phi\theta}}-\tilde{b}_{\phi r\theta})/r\big)\\
\alpha_{\phi\phi} &=& \tilde{a}_{\phi\phi}-(\underline{\tilde{b}_{\phi\phi r} + \cot \theta\, \tilde{b}_{\phi\phi\theta}})/r
\end{eqnarray}
\begin{eqnarray}
\gamma_r &=& \frac{1}{2}\big(\tilde{a}_{\phi\theta}-\tilde{a}_{\theta\phi}+(\underline{\tilde{b}_{\theta\phi r} + \cot \theta\, \tilde{b}_{\theta\phi\theta}}+\tilde{b}_{\phi r\theta})/r\big)\\
\gamma_\theta &=& \frac{1}{2}\big(\tilde{a}_{r\phi}-\tilde{a}_{\phi r}+(\tilde{b}_{\phi\theta\theta} -\underline{\tilde{b}_{r\phi r} - \cot \theta\, \tilde{b}_{r\phi\theta}})/r\big)\\
\gamma_\phi &=& \frac{1}{2}\big(\tilde{a}_{\theta r}-\tilde{a}_{r\theta}-(\tilde{b}_{rr\theta} +\tilde{b}_{\theta\theta\theta})/r\big)
\end{eqnarray}
\begin{eqnarray}
\beta_{rr} &=& -\underline{ 1}\cdot\tilde{b}_{r\phi\theta}\\
\beta_{r\theta}=\beta_{\theta r} &=&\underline{ \frac{1}{2}}(\tilde{b}_{r\phi r}-\tilde{b}_{\theta\phi\theta})\\
\beta_{r\phi}=\beta_{\phi r} &=& \frac{1}{4}(-\underline{ 2}\tilde{b}_{\phi\phi\theta}+\tilde{b}_{r r\theta}-\tilde{b}_{r \theta r})\\
\beta_{\theta\theta} &=& \underline{ 1}\cdot\tilde{b}_{\theta\phi r}\\
\beta_{\theta\phi}=\beta_{\phi \theta} &=& \frac{1}{4}(\underline{ 2}\tilde{b}_{\phi\phi r}+\tilde{b}_{\theta r\theta}-\tilde{b}_{\theta \theta r})\\
\beta_{\phi\phi} &=& \frac{1}{2}(\tilde{b}_{\phi r\theta} -\tilde{b}_{\phi\theta r})
\end{eqnarray}
\begin{eqnarray}
\delta_r &=&  \frac{1}{4}(\tilde{b}_{\theta\theta r}-\tilde{b}_{\theta r \theta} +\underline{ 2} \tilde{b}_{\phi\phi r})\\
\delta_\theta &=&  \frac{1}{4}(\tilde{b}_{rr\theta}-\tilde{b}_{r\theta r} +\underline{ 2} \tilde{b}_{\phi\phi \theta})\\
\delta_\phi &=&  \underline{ -\frac{1}{2}}(\tilde{b}_{r \phi r}+\tilde{b}_{\theta \phi\theta})
\end{eqnarray}
\begin{eqnarray}
\kappa_{irr} &=& -\tilde{b}_{irr}\\
\kappa_{ir\theta}=\kappa_{i\theta r} &=& -\frac{1}{2} (\tilde{b}_{ir\theta}+\tilde{b}_{i\theta r})\\
\kappa_{ir\phi}=\kappa_{i\phi r} &=& \underline{ 0} \\
\kappa_{i\theta\theta} &=& -\tilde{b}_{i\theta\theta}\\
\kappa_{i\theta\phi}=\kappa_{i\phi\theta} &=& \underline{0} \\
\kappa_{i\phi\phi}&=& 0
\end{eqnarray}

The results from the new definition are shown in \Fig{fig:alpC1} 
for $\aalpha$ and $\ggamma$
and in \Figu{fig:betdelkapC1} 
for the six independent components of $\bbeta$, the vector $\ddelta$
(first three columns),
and for the nine independent nonzero components
of $\kkappa$ (last three columns).
$\bbeta$, $\ddelta$, and $\kkappa$
are normalized by
$\eta_{t0}= u'_{\trms} \alpha_{\rm MLT} H_p/3$,    
where $\alpha_{\rm MLT}=5/3$ is
the mixing-length parameter and
$H_p=-1/\partial_r {\rm ln}\, p $ is the pressure scale height. 
The terms contributing to the magnetic field evolution from the 
$\delta$ (R\"adler) effect, using the new definition, are shown in \Fig{fig:deleff}.

\section{Comparison of  the turbulent transport coefficients to \cite{WRTKKB18}}

\begin{table*}[t!]
\centering
\caption{Comparison with \cite{WRTKKB18}. $q$ is
the ratio of the respective extremal value from the present study    
to that of \cite{WRTKKB18}.}
       \label{tab:coeff}
      $$
          \begin{array}{p{0.05\linewidth}c@{\hspace{3mm}}l|cc@{\hspace{3mm}}l|cc@{\hspace{3mm}}l}
            \hline
            \hline
            \noalign{\smallskip}
{\rm Coeff} & {q} & {\rm Comments} & {\rm Coeff} & {q} & {\rm Comments} & {\rm Coeff} &  {q} & {\rm Comments} \\[1mm]
            \hline
            $\alpha_{rr}$ &  0.8 &   &  \beta_{rr} & 2.4  &   & \kappa_{rrr} & 2.9  &   \\ 
            $\alpha_{r\theta}$ & 1.9   & \rm opposite~sign  &  \beta_{r\theta} & 2.5 & \rm opposite~sign &  \kappa_{rr\theta} & 2.3  &   \\[-1mm]
               & &\rm below~surface & & & \rm in~deep~CZ & & & \\
            $\alpha_{r\phi}$ & 0.7  & \rm opposite~sign~near  &  \beta_{r\phi} & 2.9  &   &  \kappa_{r\theta\theta} &  3.5 & \rm negative~near  \\[-1mm]
               & &\rm equator~in~upper~CZ & & & & & & \rm surface \\
            $\alpha_{\theta\theta}$ & 1.2 & \rm opposite~sign  &  \beta_{\theta\theta} & 4.5  & \rm weakly~negative~layer   & \kappa_{\theta rr} & 7.0  &   \\[-1mm]
               & & & & &\rm at~surface & & & \\ 
            $\alpha_{\theta\phi}$ & 1.1  &   &  \beta_{\theta\phi} & 5.0  &\rm opposite~sign  & \kappa_{\theta r\theta}  & 3.7  & \rm opposite~sign \\[-1mm] 
               & & & & &\rm near~surface & & & \rm near~surface\\
            $\alpha_{\phi\phi}$ & 0.8  &   &  \beta_{\phi\phi} &  1.1 &   & \kappa_{\theta\theta\theta} & 6.5 &  \\ 
            $\gamma_{r}$ & 1.0  &   &  \delta_{r} & 4  & \rm opposite~sign  & \kappa_{\phi rr}  & 1.1 &  \rm opposite~sign \\[-1mm] 
               & & & & &\rm at~surface & & & \\
            $\gamma_{\theta}$ & 0.6  &   &  \delta_{\theta} & 2.4  &   & \kappa_{\phi r\theta}  &  1.0 & \rm negative~layer \\[-1mm]
               & & & & & & & &\rm at~surface  \\ 
            $\gamma_{\phi}$ & 1.4  &   &  \delta_{\phi} & 3.3  &  & \kappa_{\phi\theta\theta}  & 1.4 &   \\[1mm] 
\hline
          \end{array}
          $$
\end{table*}

We summarize the ratios of the turbulent transport coefficients
from this study
and their corresponding values from
\cite{WRTKKB18}~in \Table{tab:coeff}.
Note that all coefficients except $\beta_{\phi\phi}$ and the nonvanishing 
components of $\kkappa$ are affected
by the redefinition explained in \App{sec:new_deco},
and $\beta_{rr}$ and $\beta_{\theta\theta}$
are now twice as large as with the old definition.   
The extrema of the $\beta_{ij}$ are between 2.4 and 5 times
larger than the ones in \cite{WRTKKB18}, with only $\beta_{\phi\phi}$
having the same order of magnitude, while all the 
components of $\ddelta$ are between 2.4 and 4 times larger.
The diagonal components of $\bbeta$ all show positive values
throughout the domain, except for a thin layer near the surface 
for $\beta_{\theta\theta}$.
$\beta_{r\theta}$ is positive at high latitudes and shows a sign 
reversal at the bottom of the convection zone at low latitudes.
$\beta_{\theta\phi}$ is symmetric about the equator and 
changes sign in depth.
$\beta_{r\phi}$ has a positive layer outside the tangent cylinder
and is near zero everywhere else.

$\delta_r$ changes sign at high latitudes and, with respect to 
$\delta_r$ in \cite{WRTKKB18}, has the opposite sign at low latitudes 
near the surface.
Like $\beta_{r\phi}$, $\delta_{\theta}$ has also a positive layer 
outside the tangent cylinder and two negative patches are present,
roughly at the same location as the minimum in $\Omega$.
$\delta_{\phi}$ is 1.5 times larger than by the old definition.

\begin{figure}
\begin{center}
\includegraphics[width=0.48\columnwidth]{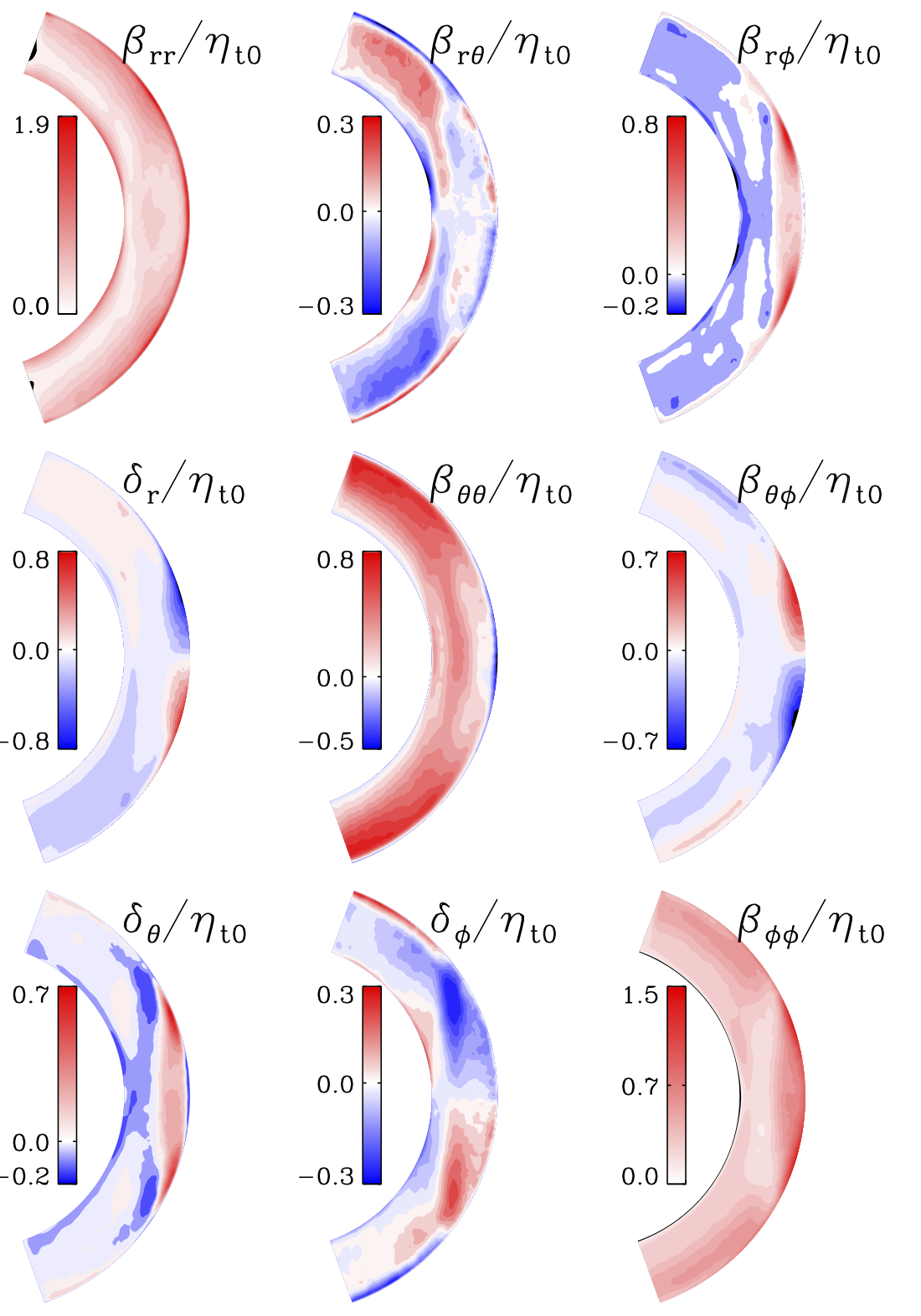}
\includegraphics[width=0.48\columnwidth]{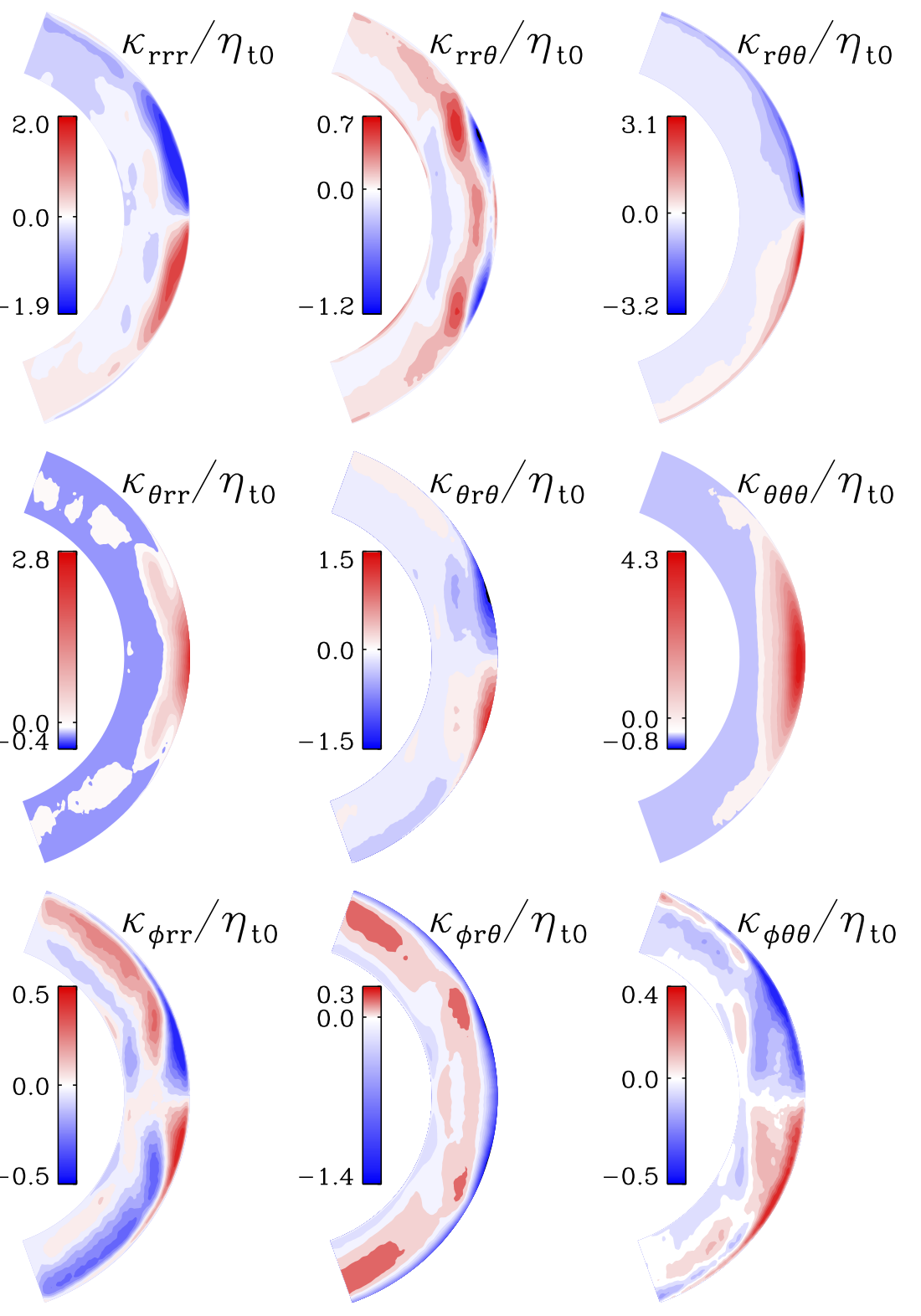}
\end{center}\caption{Independent components of time-averaged
$\bbeta$ and  $\ddelta$ 
(first three columns), and the nine
independent components of time-averaged
$\kkappa$ 
(last three columns), normalized by $\eta_{t0}= u'_{\trms} \alpha_{\rm MLT} H_p/3$,}
\end{figure}\label{fig:betdelkapC1}

\begin{figure}
\begin{center}
\includegraphics[width=0.8\columnwidth]{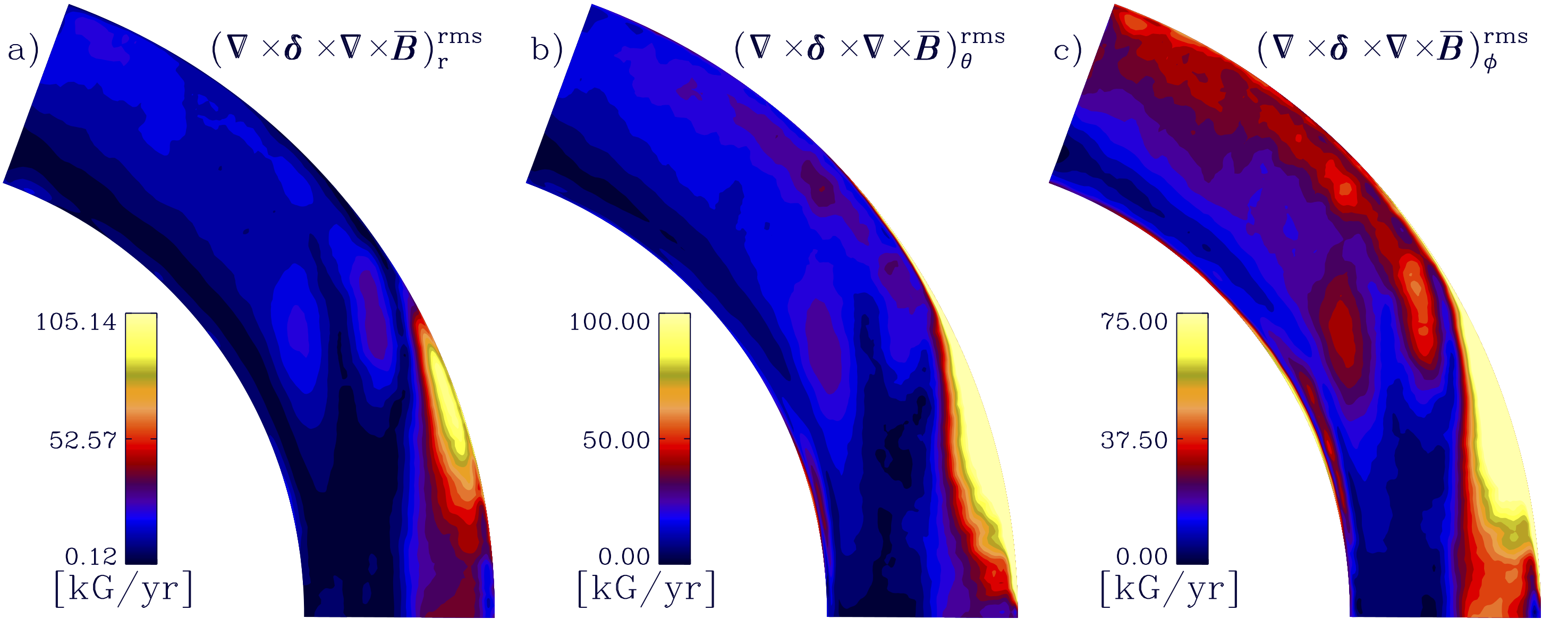}
\end{center}
\caption{Temporal rms of the components of the $\delta$
 (R\"adler) effect, $\nab\times\ddelta\times\nab\times\mBBB$,
see \Eq{eq:emf}.}
\end{figure}\label{fig:deleff}

The $\kkappa$ components look, in general, 
smoother than in \cite{WRTKKB18}.
Most of the $\kappa_{ijk}$ are now zero, leaving just nine
independent nonzero components.
$\kappa_{rrr}$, $\kappa_{rr\theta}$, $\kappa_{\theta r\theta}$
and $\kappa_{r\theta\theta}$ are roughly three times larger, 
$\kappa_{\phi\theta\theta}$ and $\kappa_{\phi r \theta}$ 
have similar magnitudes, while $\kappa_{\theta rr}$ and
$\kappa_{\theta\theta\theta}$ are 7 and 6.5 times larger 
in the current study, respectively.
$\kappa_{rr\theta}$ shows sign reversal near the surface, 
and $\kappa_{r\theta\theta}$ does not show any particular structure 
in the bulk of the convection zone, as was the case in \cite{WRTKKB18}, too.
$\kappa_{\theta rr}$ 
has strong positive values near the equator in the upper part 
of the convection zone, extending to mid-latitudes, 
while $\kappa_{\theta r\theta}$ is anti-symmetric with respect to the
equator, and has the opposite sign near the surface
with respect to \cite{WRTKKB18}.
Two sign reversals in depth are visible in $\kappa_{\phi rr}$, and also 
$\kappa_{\phi r \theta}$ shows three layers in depth: two narrow negative
ones at the top and bottom of the convection zone and a weakly positive 
one in the bulk. 

While $\boldsymbol{\alpha}$ and $\boldsymbol{\gamma}$ do not differ 
markedly between the compared runs, the other tensors show 
variations by up to a factor of seven compared to \cite{WRTKKB18}. 
Given that the roughly three-times higher Coriolis number of 
their run is virtually the only relevant difference to our present run, we have to assign 
these changes to the effect of rotational quenching \citep[see, e.g.][]{KPR94}. 
This is supported by the findings of \cite{BRK12} for rotating homogeneous 
turbulence who report on a reduction of $\boldsymbol{\beta}$ and 
$\boldsymbol{\delta}$ by a factor of approximately three when $\Co$ is increased 
from two to eight, with an even stronger reduction in $\boldsymbol{\kappa}$.

\section{Reconstruction of the turbulent electromotive force}
\label{sec:EMFrec}

We show in \Figu{fig:rec} the turbulent EMF,
computed directly via $\overline{\uuu'\times\bbb'}$  and its
reconstruction using \Eq{eq:emf}
with the time-averaged transport coefficients
and the full $\mBBB$
during roughly five typical dynamo cycles.
In the reconstructed and directly computed EMFs, we have filtered 
out the time average and all
time-scales shorter than one year 
to highlight the oscillating pattern.
The spatial and temporal structures of all components of the reconstructed EMF
match the measured ones reasonably well.
In \cite{WRTKKB18}, a good match was found in the mid and high latitudes, 
while the near-equator behaviour was captured less accurately.
However, the time average was not removed there.
Now we find good correspondence also at the equatorial regions. 
As in \cite{WRTKKB18}, the magnitudes of the reconstructed EMF components tend 
to be overestimated. 
Here, this effect is most pronounced for the azimuthal component
of the EMF, which is 
by a factor of 2.5 larger than the measured one. 
This can be interpreted as a consequence of non-localities in turbulent convection, and
calls for the application of scale dependent test fields to the problem.

\begin{figure}
\begin{center}
\includegraphics[width=\columnwidth]{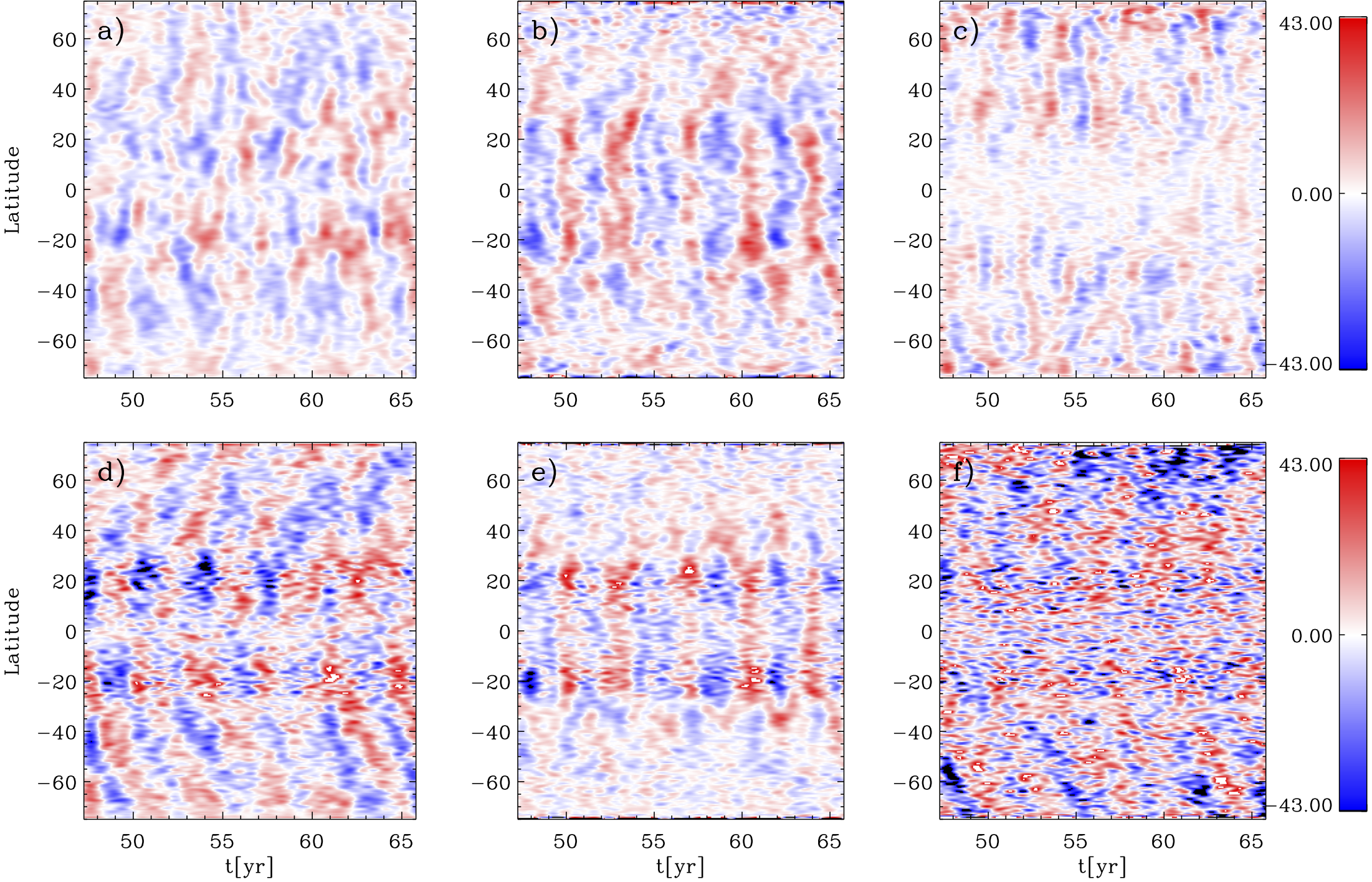}
\end{center}\caption{
Radial, latitudinal, and longitudinal components of the
turbulent electromotive force computed directly using $\overline{\uuu'\times\bbb'}$
(panels a-c) and of its reconstruction using \Eq{eq:emf} (panels d-f) near 
the surface ($r=0.98\,R$).
Time averages and time-scales shorter than one year
have been filtered out.
The longitudinal components (panels c) and f)) have been
multiplied by a factor 2.5.}

\end{figure}\label{fig:rec}

\acknowledgments
We acknowledge the HPC-EUROPA3 project (INFRAIA-2016-1-730897),
supported by the EC Research Innovation Action under the H2020 Programme.
M.V., M.J.K., P.J.K., and M.R. acknowledge the support of 
the Academy of Finland ReSoLVE Centre of Excellence (grant No.
307411).
M.V. acknowledges being enrolled in the International Max Planck Research School 
for Solar System Science at the University of G\"ottingen (IMPRS).
J.W. acknowledges funding by the Max-Planck/Princeton Center for
Plasma Physics.
P.J.K. acknowledges support from DFG Heisenberg grant (No.\ KA 4825/1-1).
This project has received funding from the European Research Council (ERC) 
under the European Union's Horizon 2020 research and innovation 
programme (project "UniSDyn", grant agreement n:o 818665).
We acknowledge also support from the supercomputers at
GWDG, at RZG in Garching, and in the
facilities hosted by the CSC---IT
Center for Science in Espoo, Finland, which are financed by the
Finnish ministry of education.

\bibliographystyle{apj}
\bibliography{bib}

\end{document}